**Emergent Electronic Bond-Order Wave in a Quasi-One-Dimensional Chain**

Honghao Wang[1,2,] Tristan R. Cao[1], Pedro Schlottmann[3], and Gang Cao[1,2*]

[1]Department of Physics, University of Colorado at Boulder, Boulder, CO 80309, USA

[2]Materials Science and Engineering, University of Colorado at Boulder, Boulder, CO 80309, USA

[3]Department of Physics, Florida State University, Tallahassee, FL 32306, USA

Collective electronic order is notoriously difficult to stabilize in one dimension, where strong fluctuations suppress symmetry-breaking states. Here we report evidence for an unconventional electronic bond-order wave (BOW) in the quasi-one-dimensional $Ba_9Rh_8O_{24}$, marked by a striking inversion of inequivalent Rh-Rh bonds across a bulk thermodynamic transition at $T_A \approx 180$ K. Unlike conventional Peierls dimerization, the short and long Rh–Rh bonds exchange identities, revealing an unprecedented reorganization of the bond hierarchy. The transition is remarkably insensitive to magnetic fields up to 14 T, while thermal conductivity exhibits an unusual linear temperature dependence that is dramatically altered by only 2% La substitution. Concurrently, the dielectric response undergoes a pronounced reorganization, and nonlinear electrical transport reveals frequency-dependent hysteresis and remanent voltage, signaling unusual collective dynamics and electrical memory. These findings point to an unconventional bond-centered electronic instability beyond conventional Peierls physics, suggesting a new paradigm for low-dimensional quantum matter in which electronic bonds become collective degrees of freedom linking structural order, lattice dynamics, and electrically reconfigurable functionality.

*gang.cao@colorado.edu

One-dimensional (1D) quantum systems occupy a unique position in condensed matter physics. Strong fluctuations, enhanced electronic correlations, and reduced dimensionality give rise to a wide variety of exotic phenomena, such as Luttinger liquids [1], spin-charge separation [2], Peierls instabilities [3], charge-density waves [3], and topological edge states [4]. Yet these same fluctuations also impede the formation of conventional long-range order, as the Mermin-Wagner theorem indicates. Indeed, reduced dimensionality strongly enhances quantum fluctuations, making the discovery of new mechanisms that stabilize collective quantum states one of the central challenges in condensed matter physics [5-8].

In recent years, it has become increasingly clear that interacting electrons can self-organize into collective states extending far beyond conventional spin, charge, and orbital order, such as hidden order [9,10], electronic nematicity [11], multipolar order [12], loop-current order [13,14], and other intertwined electronic phases [8]. Among these unconventional states, bond-centered order is particularly intriguing because the collective degree of freedom resides not primarily on an atomic site, but on the ***electronic bond*** connecting neighboring sites [6,15,16]. A long-sought-after ***bond-order wave (BOW)***, characterized by a spatial modulation of the electronic kinetic energy on neighboring bonds, was theoretically predicted as a distinct correlated ground state of the 1D extended Hubbard model that features the Hubbard interaction U and the nearest-neighbor charge interaction V [15]. Here, 2V = U is the boundary between the charge density wave (CDW) and spin density wave (SDW) phases; the BOW, extremely sensitive to small perturbations, can only exist very close to this boundary, before the CDW-SDW transition becomes first order [15]. The BOW was subsequently established by quantum Monte Carlo calculations [16-18], making the bond, rather than the atomic site, the fundamental electronic degree of freedom. In recent years,

BOW states have been explored across various material platforms [19-23], yet ***direct experimental realizations of equilibrium electronic BOW order in quasi-1D materials remain elusive.***

Here we report evidence for precisely such a state: an emergent electronic BOW in the quasi-1D chain $Ba_9Rh_8O_{24}$ (**Fig. 1**), uncovering an unusual realization of bond-centered electronic order in a bulk single-crystal material. At the heart is a striking bond inversion: Inequivalent short ($d_S$) and long ($d_L$) Rh-Rh bonds persist on both sides of a transition at $T_A \approx 180$ K, yet exchange their identities, so that short bonds become long and long bonds become short (**Figs. 1c-1e**). This behavior contrasts sharply with conventional Peierls dimerization, which drives short bonds shorter and long bonds longer [24]. Remarkably, the inversion coincides with a distinct lattice anomaly (**Fig. 1a**), a bulk heat-capacity transition (**Fig. 2a**), and unconventional thermal conductivity (**Fig. 2b**). It is further accompanied by a profound reorganization of the dielectric response, with strongly suppressed dielectric susceptibility and loss (**Figs. 3a-3b**), and strongly nonlinear transport exhibiting frequency-dependent hysteresis and remanent voltage (**Fig. 4**). All in all, these structural, thermodynamic, transport, dielectric, and electrical signatures provide mutually reinforcing evidence for the existence of a BOW. We propose that fluctuating bond configurations above $T_A$ develop into an emergent electronic BOW below $T_A$, progressively establishing a more rigid yet electrically reconfigurable state, fundamentally distinct from conventional Peierls-type ordering [3, 24].

It is worth emphasizing that $Ba_9Rh_8O_{24}$ is a *4d*-electron system in which spin-orbit interaction (SOI ~ 0.16 eV) constitutes an important energy scale [25, 26]. More generally**,** in *4d*- and *5d*-transition-metal systems containing strong coupled clusters such as trimers, the electronic states are governed not by isolated transition-metal ions but by strongly hybridized molecular-like building blocks [27, 28]. The competition among metal-metal covalency, inter-unit coupling, electron correlations, lattice distortions and SOI can entangle spin, orbital, and kinetic degrees of freedom, generating a rich landscape of competing states in which subtle changes in bond geometry can fundamentally reorganize the ground state [29-32]. Such a delicate competition provides fertile ground for collective phenomena rarely accessible in conventional low-dimensional systems [27-32]. $Ba_9Rh_8O_{24}$, with its quasi-1D chains and sizable SOI, provides a compelling example.

***Crystal structure and bond inversion.*** $Ba_9Rh_8O_{24}$ crystallizes in the rhombohedral space group $R\bar{3}c$ and belongs to the family of hexagonal perovskite-derived rhodates composed of quasi-1D chains extending along the crystallographic *c* axis [33]. The crystal structure consists of infinite Rh-O chains formed by seven face-sharing $RhO_6$ octahedra separated by a single face-sharing $RhO_6$ trigonal prism, while neighboring chains are well isolated by Ba ions, resulting in a highly 1D electronic environment (**Fig.1b**). There are four inequivalent Rh-Rh bonds within each Rh chain (**Fig.1c**). At room temperature, the Rh-Rh bonds are already inequivalent, reflecting alternating short ($d_S$) and long ($d_L$) Rh-Rh separations rather than a uniform chain [33]. Our temperature-dependent single-crystal X-ray diffraction shows pronounced anomalies near a transition $T_A \approx 180$ K in the lattice parameter *a*-, *c*-axis and unit-cell volume *V* (**Fig.1a**). Strikingly, it reveals that the identities of the short and long bonds interchange across $T_A$ (**Figs.1c-1d**), indicating that the Rh-Rh bonds participate directly in a collective electronic instability, ultimately giving rise to the emergent BOW. More experimental details can be found in Supplemental Information [34].

In addition, charge neutrality requires mixed Rh valence states in $Ba_9Rh_8O_{24}$, corresponding nominally to six $Rh^{4+}(4d^5)$ and two $Rh^{3+}(4d^6)$ ions per formula unit, indicating that $Rh^{4+}$ constitutes the majority electronic species. The previous crystallographic study found that bond-valence analysis could not uniquely assign these oxidation states to specific crystallographic sites [33], suggesting that the valence distribution is not simply localized on individual Rh ions. This intrinsic mixed-valence character naturally promotes electronic hybridization and bond-centered charge redistribution along the face-sharing Rh chains. The majority $Rh^{4+}$ ions are expected to participate collectively through direct Rh-Rh interactions.

***Heat capacity, thermal conductivity, magnetization, and dielectric response.*** The bond inversion is accompanied by a well-defined thermodynamic phase transition (**Fig.2)**. The heat capacity, C(T), exhibits a pronounced anomaly at $T_A \approx 180$ K, establishing that the transition is an undisputable bulk phase transition (right axis in **Figs.2a-2b**). Remarkably, this transition is insensitive to magnetic fields, H, at least up to 14 T, signaling a non-magnetic nature of the ordering process (**Fig.2a** and SFig.1 [34]). The *c*-axis resistivity, $\rho_c$, also shows a broad slope change near $T_A$ (left axis in **Fig.2a**) (Note that the material becomes too resistive for resistivity measurements below 110 K).

Furthermore, the thermal conductivity along the *c* axis, $\kappa_c$, offers important insights into the transition (**Fig.2b**). Strikingly, $\kappa_c(T)$ increases linearly between 30 and 180 K, without the conventional high-T downturn associated with Umklapp scattering. Above $T_A \approx 180$ K, $\kappa_c(T)$ deviates upward from the linear behavior, coincident with the anomaly in CT (right scale, **Fig.2b**) and Rh-Rh bond inversion (**Fig.1c**). Moreover, $\kappa_c(T)$ also exhibits weak H-dependence up to 14 T, consistent with the field insensitivity of the transition. These observations suggest unconventional phonon scattering associated with bond fluctuations and their reorganization across the BOW transition. Below 30 K, $\kappa_c(T)$ decreases rapidly upon cooling, consistent with the diminishing phonon population in the low-T Debye regime (Debye temperature $\theta_D = 274$ K).

To further probe the unusual thermal transport, we also measured $\kappa_c(T)$ of a 2% La-substituted sample, $(Ba_{1-x}La_x)_9Rh_8O_{24}$ with x = 0.02 (**Fig.2c**). Importantly, the modest La substitution readily disrupts the linear $\kappa_c(T)$ observed between 30 and 180 K in the pristine compound, producing pronounced curvature and substantially enhanced $\kappa_c(T)$ at higher T despite the introduction of chemical disorder. In sharp contrast, C(T) for the doped sample retains a comparatively smooth T-dependence (SFigs.1-2). Since $\kappa \approx 1/3\, Cvl$ ($v$ = velocity of heat carriers and $l$ = mean free path), the contrasting behavior of κ points toward a high susceptibility to changes in $l$ or unconventional phonon scattering associated with fluctuations of competing Rh-Rh bond configurations.

Additionally, the *c*-axis magnetization, $M_c$, exhibits a distinct valley near $T_A$, without the presence of a Curie-Weiss behavior above $T_A$ (**Fig.2d**). In contrast, this anomaly is absent in the *a*-axis magnetization, $M_a$ (left scale in **Fig.2d**). Since SOI entangles spin and orbital degrees of freedom, reorganizing the Rh-Rh bonds (**Figs.1c-1e**) is expected to also modify the magnetization. As such, the modest magnetic anomaly near $T_A$ in $M_c$ is compatible with an SOI-assisted bond-order transition rather than a conventional spin transition. Indeed, the isothermal magnetization, M(H), for both the *a* and *c* axis decreases rapidly upon warming, and becomes vanishingly small at 100 K, which is still well below $T_A$ (SFig. 3).

In addition, the entropy change associated with the transition at $T_A \approx 180$ K (**Fig.2a**) is estimated to be small, $\Delta S_A \approx 1.6$ J/ mole K. The value corresponds to only ~ 28% of Rln2 (= Rln (2S+1), where R = gas constant) and less than 5% of the total spin entropy expected for the six nominal $Rh^{4+}$ (S = ½) ions per formula unit. The small $\Delta S_A$ confirms that the transition does not correspond to conventional magnetic ordering of localized moments, but rather it is more consistent with a reorganization of electronic bond degrees of freedom at $T_A$ where only a fraction of the available electronic entropy is released. The small $\Delta S_A$ also implies that substantial bond correlations develop well above $T_A$ in a dynamic bond liquid before condensing into the BOW. Collectively, these results reveal a remarkably robust, predominantly nonmagnetic bulk instability associated with the unprecedented Rh-Rh bond inversion, yet with thermal transport extraordinarily sensitive to only 2% La substitution, consistent with the theoretical anticipation [15]. The convergence of structural, thermodynamic, thermal, and magnetic evidence strongly supports an unconventional bond-centered ordering transition.

Equally important, the *c*-axis dielectric response undergoes a pronounced reorganization across $T_A$, as shown in **Fig. 3a**, providing compelling evidence that the transition involves a strongly polarizable electronic degree of freedom. The real dielectric constant, $\varepsilon_c$', which measures the non-dissipative dielectric response, develops a pronounced anomaly near $T_A$ (**Fig.3a**). It is nearly frequency (*f*) independent above and near $T_A$, while only weak *f*-dependence emerges at T < 100 K (**Fig. 3b**). Such behavior points to an intrinsic electronic polarizability rather than a conventional relaxor response. The imaginary component, $\varepsilon_c$'', associated with dielectric dissipation, is large above $T_A$ but strongly suppressed below $T_A$ (**Fig. 3a**). The significantly reduced $\varepsilon_c$' and $\varepsilon_c$'' below $T_A$ (**Figs. 3a-3b**) signals a considerable suppression of dissipative electronic fluctuations and the emergence of the rigid bond-ordered state.

In **Figs. 3c-3d**, both $\varepsilon_c$' and $\varepsilon_c$'' exhibit a pronounced *f*-dependence below $10^3$ Hz, indicating a slow dielectric relaxation associated with fluctuating electronic degrees of freedom. Note that for low *f*, $\varepsilon_c' \propto 1/f^2$ and $\varepsilon_c'' \propto 1/f$. Above $10^3$ Hz, however, $\varepsilon_c$' approaches a broad, nearly *f*-independent plateau (**Fig.3c**). The simultaneous collapse of dielectric response and loss indicates a progressive freezing of fluctuating electronic degrees of freedom upon cooling, consistent with the transition from a dynamic bond state at $T > T_A$ into a rigid BOW at $T < T_A$.

***Nonlinear transport and nonvolatile memristive switching.*** The formation of the rigid BOW creates a collectively pinned yet electrically reconfigurable state and is manifested by strongly nonlinear I-V characteristics, threshold behavior, and nonvolatile memristive switching (**Fig. 4)**. At 0.3 Hz, the I-V characteristic evolves from linear at 300 K to conventional counterclockwise hysteresis at 200 K and, upon cooling to 150 K, to an abrupt switching characterized by a depinning threshold, $V_{DP}$ (**Fig. 4a**).

The nonlinear response persists over a broad frequency range and develops particularly striking ***clockwise loops***, for which a finite remanent voltage, $V_m$, remains when the current returns to zero (**Fig. 4b**). Furthermore, as the current is reduced in the return branch, the voltage remains nearly unchanged or even slightly increases, most clearly at 120 K, before reaching the finite $V_m$ at I = 0; for example, at 120 K, $V_m \simeq 4.2$ V at I = 0 (**Fig. 4b** and SFig.4). This behavior demonstrates that the electrical response is not determined by the instantaneous current alone but retains a strong memory of the current-driven state established on the forward sweep. We attribute this behavior to current-driven reorganization of the bond configuration: Once driven into a new configuration, the bond network cannot fully retract when the current is removed, leaving a metastable state with finite $V_m$. Indeed, the slight increase in V with decreasing I further suggests an inductive-like response arising from the delayed reorganization of the BOW (SFig.4).

A weak clockwise response survives even at 200 K, above $T_A$, indicating that current-reconfigurable bond dynamics already exists at $T > T_A$ but with vanishing $V_m$ (**Fig.4b**). Note that all displayed I-V curves comprise two consecutive sweeps; hence, the subsequent sweep begins from the metastable configuration created by the preceding sweep, accounting for its finite, rather than zero, starting point (**Fig. 4b**).

The contrasting *f*-dependence across $T_A$ further distinguishes the rigid BOW from the dynamic bond liquid. Below $T_A$, clockwise loops persist throughout $0.3 \leq f \leq 97.6$ Hz and become increasingly pronounced with increasing *f* (**Fig. 4c**), indicating that the current-reorganized BOW remains metastable even at the longest driving timescale. Remarkably, the metastable configuration renders finite $V_m$ (**Fig. 4b** and SFig.4). Above $T_A$, by contrast, conventional counterclockwise loops dominate at low frequencies (0.3 - 3 Hz), while clockwise loops emerge only for $f > 3$ Hz; however, $V_m$ at I = 0 either vanishes or becomes vanishingly small (**Fig. 4d**). This crossover at $T > T_A$ reveals the sluggish yet relaxational character of the dynamic bond liquid. Nevertheless, the contrasting behavior displayed in **Figs.4c** and **4d** provides a dynamical signature of the rigidity acquired upon formation of the BOW below $T_A$.

The nonlinear transport is strikingly reminiscent of the nonvolatile memristive response [35] associated with chiral orbital currents (COC) [14], where an applied current reorganizes a COC configuration into metastable states that persist after removal of the drive [35]. The close phenomenological correspondence between COC and BOW points to a broader principle: Nonvolatile memristive switching may serve as a ***fingerprint*** of electrically reconfigurable collective order, arising from current-driven reorganization of COC in one case [35, 36] and of the BOW in the present system.

**Fig. 5** summarizes the proposed evolution of the electronic bond state. At $T > T_A$, inequivalent short ($d_S$) and long ($d_L$) Rh-Rh bonds are already present, but their configurations remain strongly fluctuating, defining a ***dynamic bond liquid*** (**Fig.5a**). On cooling toward $T_A \approx 180$ K, the lattice undergoes a collective reorganization marked by the interchange of the short ($d_S$) and long ($d_L$) Rh-Rh bonds or ***bond inversion*** (**Fig. 5b**), alongside the concurrent anomalies in heat capacity (**Figs. 2a** and **5a**), resistivity (**Fig. 2a**), thermal conductivity (**Figs.2b-2c**), magnetization (**Fig. 2d**), dielectric response (**Fig. 3a**). The BOW emerges in this regime. Further cooling drives the system into a ***rigid BOW*** with a strongly suppressed $\varepsilon_c$' and vanishingly small $\varepsilon_c$'' (**Figs. 3a** and **5c**) and a strongly nonlinear, history-dependent electrical state (**Figs. 4** and **5c**). The resulting progression - ***Dynamic bond liquid* → *Emergent BOW* → *Rigid BOW*** - provides a unified framework connecting the structural, thermodynamic, dielectric, and nonlinear transport properties of this quasi-1D system.

The striking bond inversion raises a fundamental question: Why does $Ba_9Rh_8O_{24}$ reverse the Rh-Rh bond hierarchy rather than undergo conventional Peierls dimerization? We propose that its cluster-based electronic structure, intermediate Rh valence, and strong covalency favor competing, nearly degenerate bond configurations, whose relative stability may be further influenced by SOI. Unlike conventional dimerization [24], the inversion suggests that electronic energy gained through bond reorganization overcomes the elastic cost. The robust, field-insensitive thermodynamic transition, unconventional thermal conductivity highly sensitive to slight doping, and striking nonlinear electrical memory collectively point to an electronically driven bond-ordering instability. We propose that fluctuating bond configurations above $T_A$ condense into an inverted bond hierarchy below $T_A$, giving rise to an emergent BOW. These findings reveal a new paradigm for low-dimensional quantum matter: electronic bonds themselves can become collective degrees of freedom, linking unconventional order, lattice dynamics, and electrically reconfigurable functionality.

**Acknowledgement**

This work was supported by the US Department of Energy via Award DE-SC0025273. We acknowledge discussions with Joey Augustine, Tiantian Wang and Haley Cunningham.

## Figure captions

**Fig. 1. Crystal structure and temperature-driven Rh–Rh bond inversion in $Ba_9Rh_8O_{24}$.** **a,** Temperature dependence of the lattice parameter *a* and *c* axis and unit-cell volume *V*. **b,** Crystal structure viewed perpendicular and parallel to the *c* axis. The infinite Rh-O chains are purple, and Ba, Rh and O ions are green, purple and red spheres. **c,** Temperature evolution of four inequivalent intrachain Rh-Rh distances. **d, e,** Enlarged views of the Rh chain at 100 and 250 K, respectively, illustrating the reversal of the four inequivalent Rh-Rh bond lengths across $T_A$. **f,** Extended quasi-1D Rh chain emphasizing the structural motif along the *c* axis.

**Fig. 2. Heat capacity, magnetic and transport signatures of the bond-ordering transition.** **a,** Temperature dependence of the *c*-axis resistivity $\rho_c$ (left axis) at 0.01 mA and 0 T and heat capacity C(T) measured at 0 and 14 T (right axis). **b,** Temperature dependence of the *c*-axis thermal conductivity $\kappa_c(T)$ at 0 and 14 T (left axis), compared with C(T) (right axis). The dashed green line highlights the linear increase in $\kappa_c$ between 30 and 180 K, contrasting with conventional Umklapp-dominated phonon transport. Note near $T_A$ = 180 K, $\kappa_c$ deviates from the linear T dependence. **c.** Comparison of $\kappa_c(T)$ between pristine and 2% La-substituted samples, revealing pronounced curvature and enhanced high-T $\kappa_c(T)$ upon modest substitution. The dashed line represents the linear guide for the pristine compound. **d**, Temperature dependence of the *a*- and *c*-axis magnetization $M_a$ and $M_c$ at $\mu_o H$ = 0.5 T, compared with C(T) (right axis).

**Fig. 3 Reorganization and freezing of the dielectric response across the bond-ordering regime.** **a,** Temperature dependence of the real $\varepsilon_c$' and imaginary $\varepsilon_c$'' components of the *c*-axis dielectric constant measured at 50 kHz. **b,** $\varepsilon_c$' measured at 25, 50, and 75 kHz, showing the insensitivity of $T_A$ to *f*. **c, d,** *f*-dependence of $\varepsilon_c$' and $\varepsilon_c$'', respectively, at selected temperatures.

**Fig. 4. Nonlinear transport, dynamical hysteresis, and electrical reconfigurability of the bond-ordered state. a,** Temperature-dependent I-V characteristics measured at *f* = 0.3 Hz for 150, 200 and 300 K. Note a characteristic depinning voltage $V_{DP}$ (thin arrow) at 150 K. **b,** I-V characteristics measured at *f* = 97.6 Hz for 120, 150, 200 and 300 K. Note a finite voltage $V_m$ (thick arrows) as I = 0 in the return branch. **c, d,** *f*-dependent I-V characteristics at 150 K and 200 K, respectively. Note the occurrence of counter-clockwise (thicker curves) and clockwise (thinner curves) loops, and that all curves represent two consecutive voltage sweeps, thus a subsequent sweep can begin at finite V. Note that in **c**, the V-axis starts at V = 1 V for clarity and that in **d**, $V_m$ = 0 for all *f* except for 48.8 Hz.

**Fig.5. Emergence, freezing, and electrical reconfiguration of electronic bond order in $Ba_9Rh_8O_{24}$. a,** Schematic summary of the proposed transition of the bond state across $T_A$. At T > $T_A$, inequivalent long (blue $d_L$) and short (red $d_S$) Rh-Rh bonds coexist while undergoing strong fluctuations, defining a dynamic bond liquid. Note that four inequivalent Rh-Rh bonds, two long and two short, are present in the actual structure (Fig. 1); for clarity, only one representative long bond (blue $d_L$) and one representative short bond (red $d_S$) are illustrated here. Approaching $T_A$, bond correlations develop into an emergent BOW, accompanied by a heat capacity anomaly. Below $T_A$, the bond inversion enables a rigid BOW. Note that the blue $d_L$ and red $d_S$ interchange their identities. **b,** The bond inversion across $T_A$. **c,** Below $T_A$, both $\varepsilon_c$' and $\varepsilon_c$'' are strongly suppressed, indicating the freezing of polarizable bond fluctuations and the presence of a rigid BOW, enabling nonvolatile memristive switching.

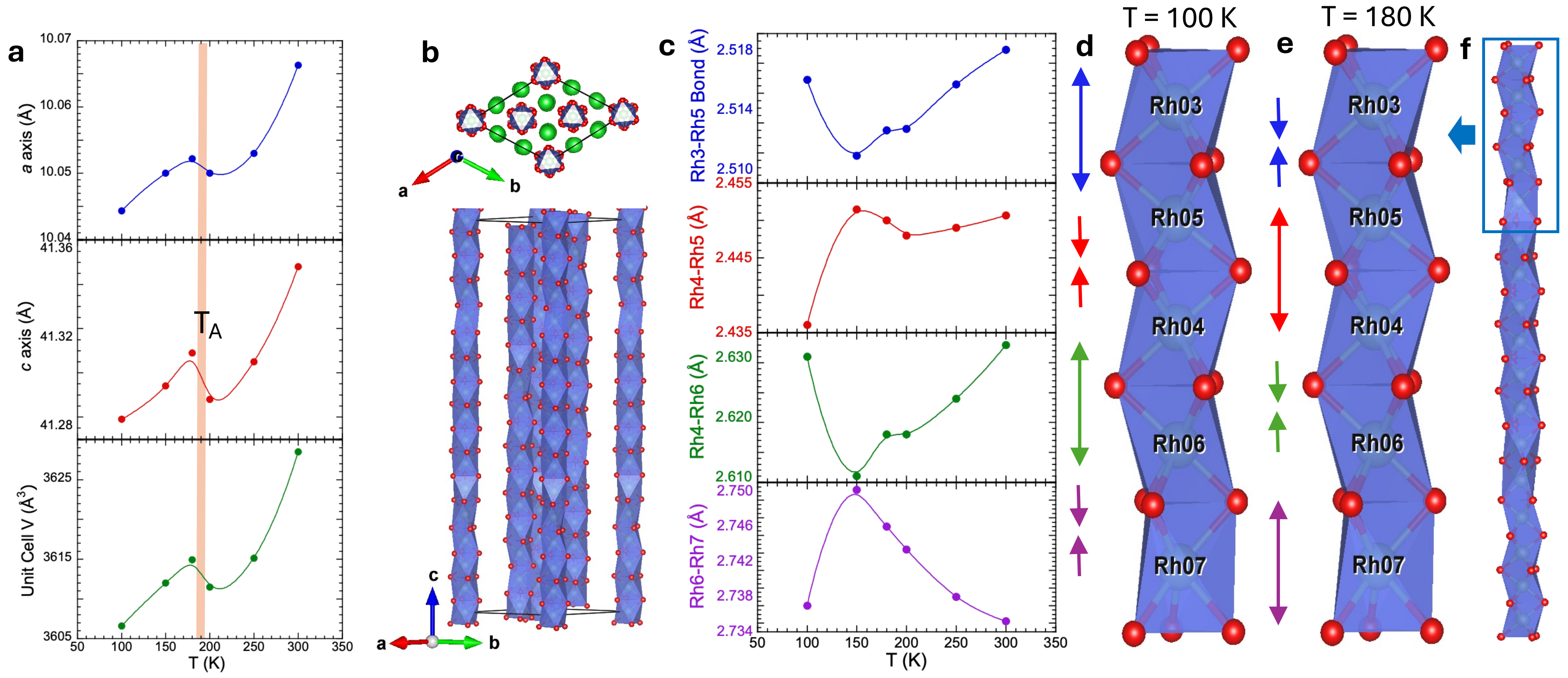

a
b
c
d
e
f
a axis (Å)
c axis (Å)
Unit Cell V (Å3)
T (K)
TA
Rh3-Rh5 Bond (Å)
Rh4-Rh5 (Å)
Rh4-Rh6 (Å)
Rh6-Rh7 (Å)
T = 100 K
T = 180 K
Rh03
Rh05
Rh04
Rh06
Rh07

Figure 1

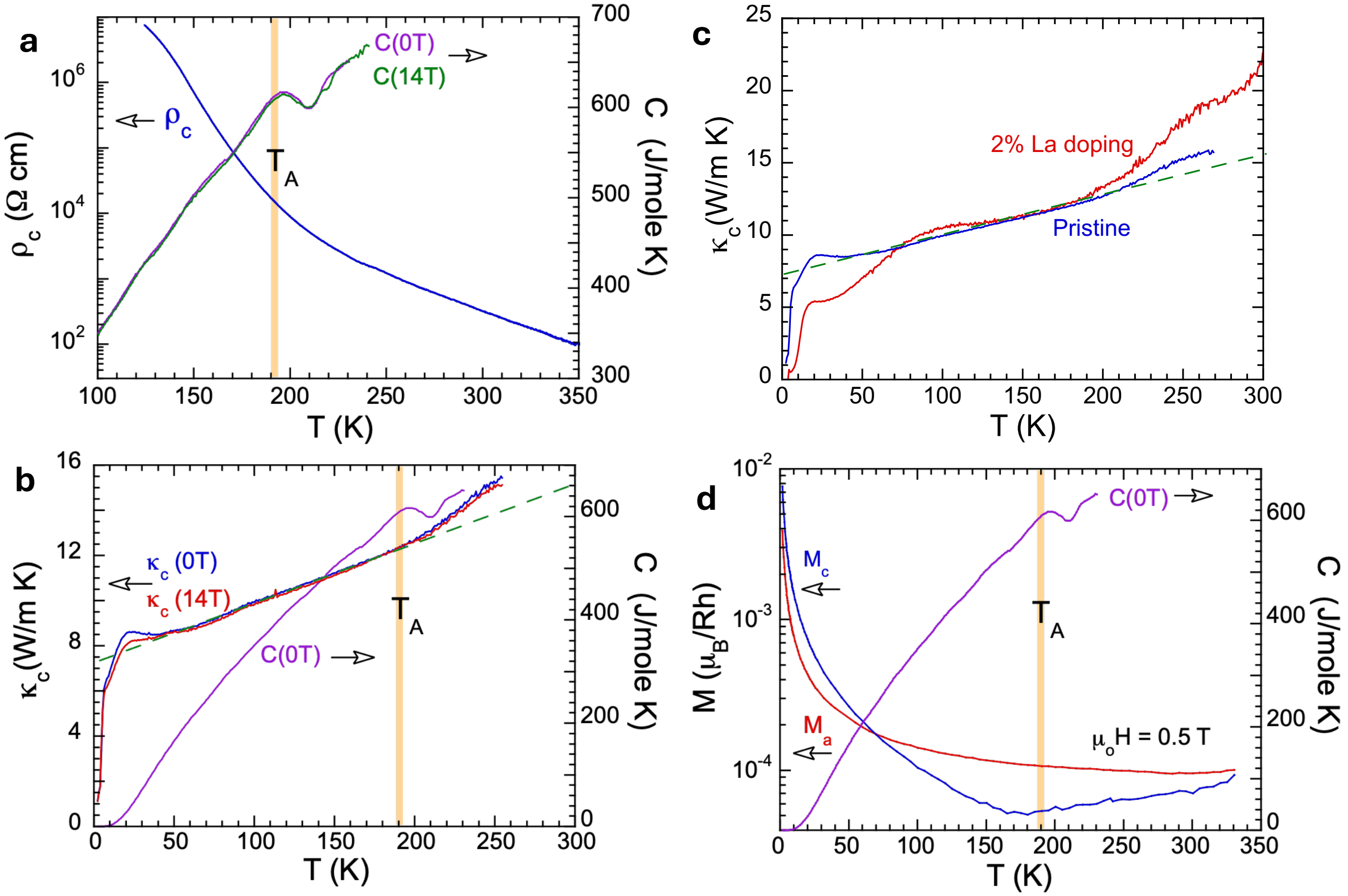
a
b
c
d
ρc (Ω cm)
C (J/mole K)
T (K)
C(0T)
C(14T)
ρc
TA
κc(W/m K)
κc (0T)
κc (14T)
2% La doping
Pristine
M (μB/Rh)
Mc
Ma
μoH = 0.5 T

Figure 2

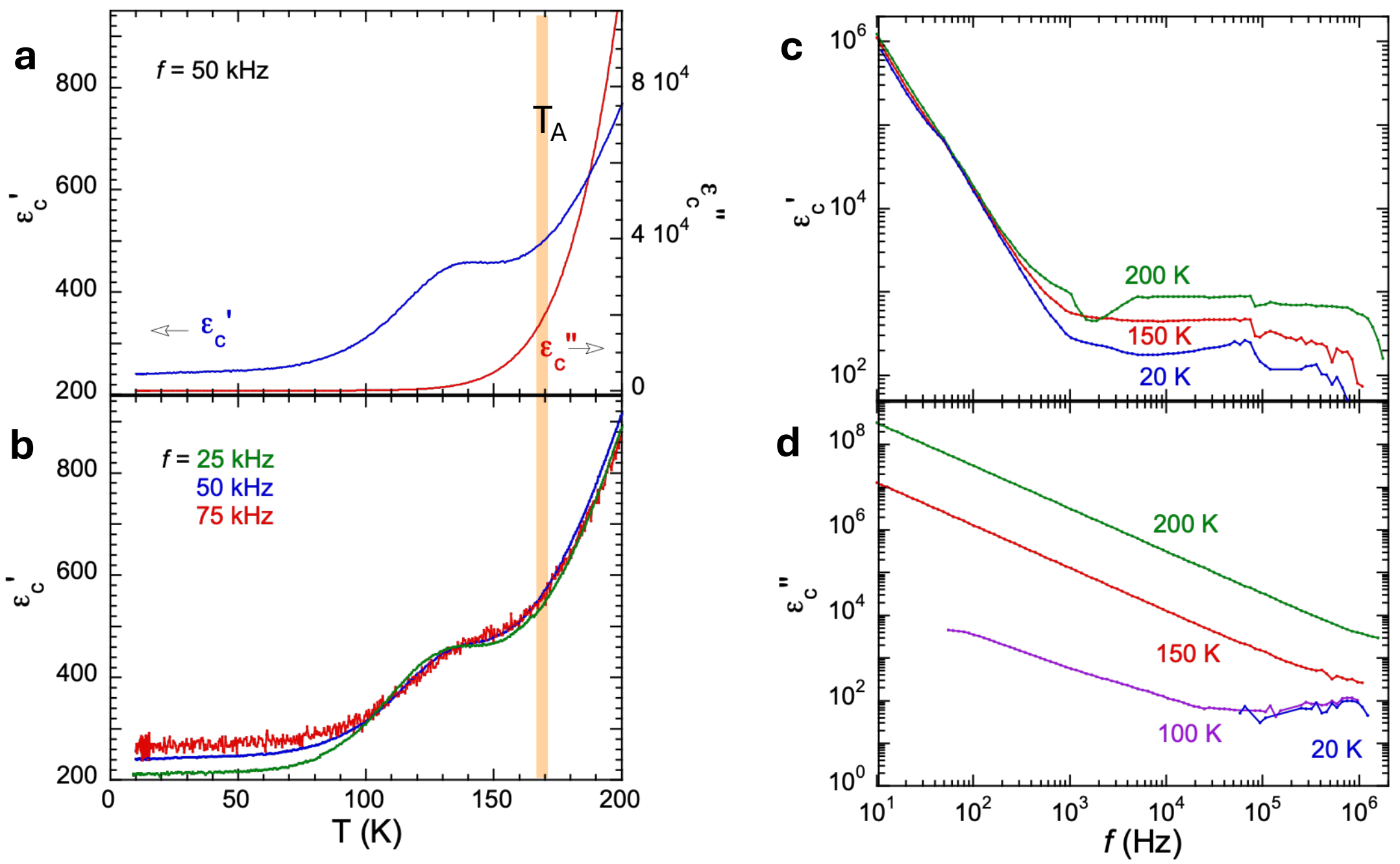

a
f = 50 kHz
T_A
ε_c'
ε_c"
b
f = 25 kHz
50 kHz
75 kHz
T (K)
c
200 K
150 K
20 K
d
200 K
150 K
100 K
20 K
f (Hz)

Figure 3

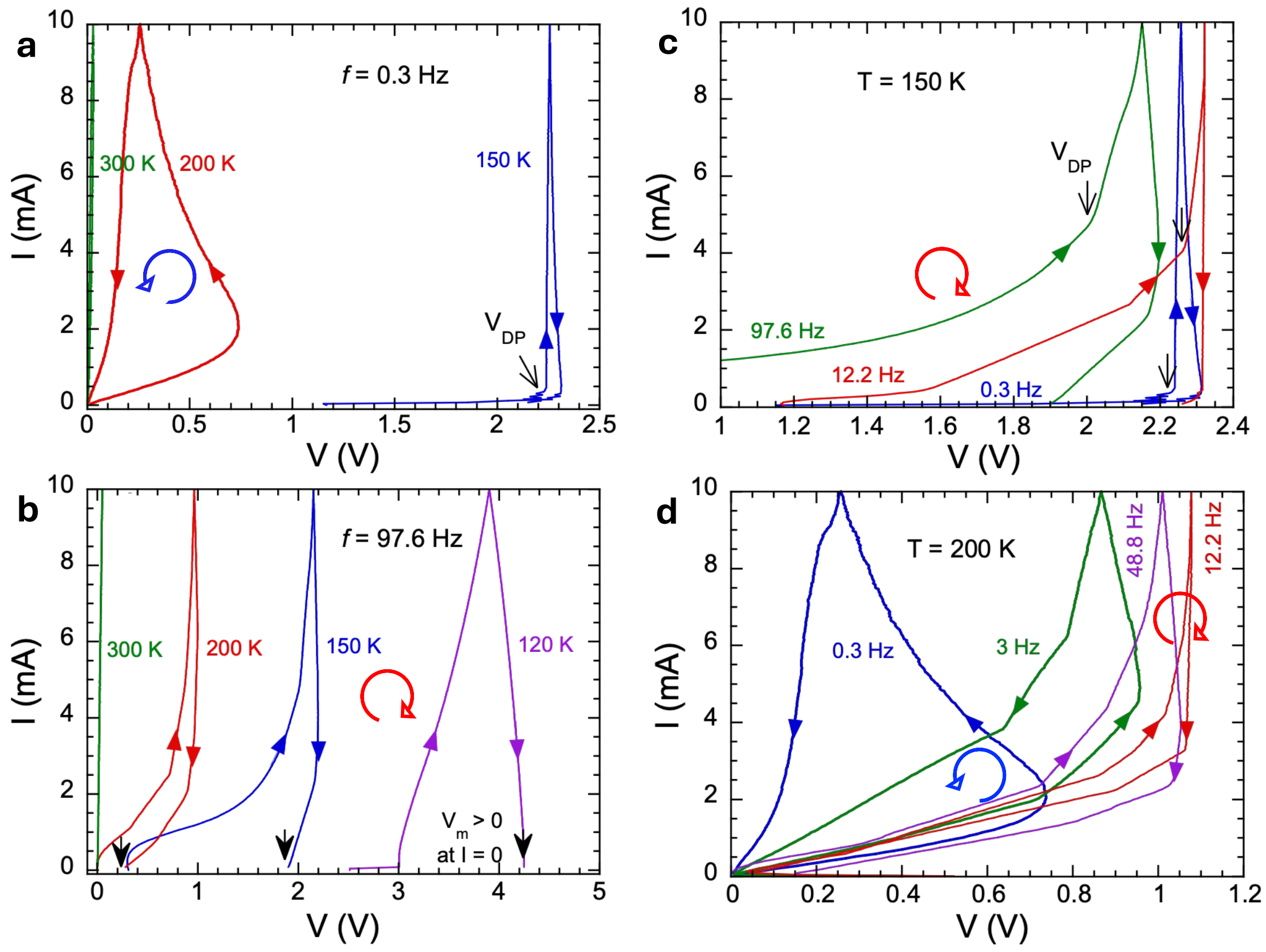

a
f = 0.3 Hz
300 K
200 K
150 K
V_DP
I (mA)
V (V)
b
f = 97.6 Hz
300 K
200 K
150 K
120 K
V_m > 0
at I = 0
I (mA)
V (V)
c
T = 150 K
V_DP
97.6 Hz
12.2 Hz
0.3 Hz
I (mA)
V (V)
d
T = 200 K
0.3 Hz
3 Hz
48.8 Hz
12.2 Hz
I (mA)
V (V)

Figure 4

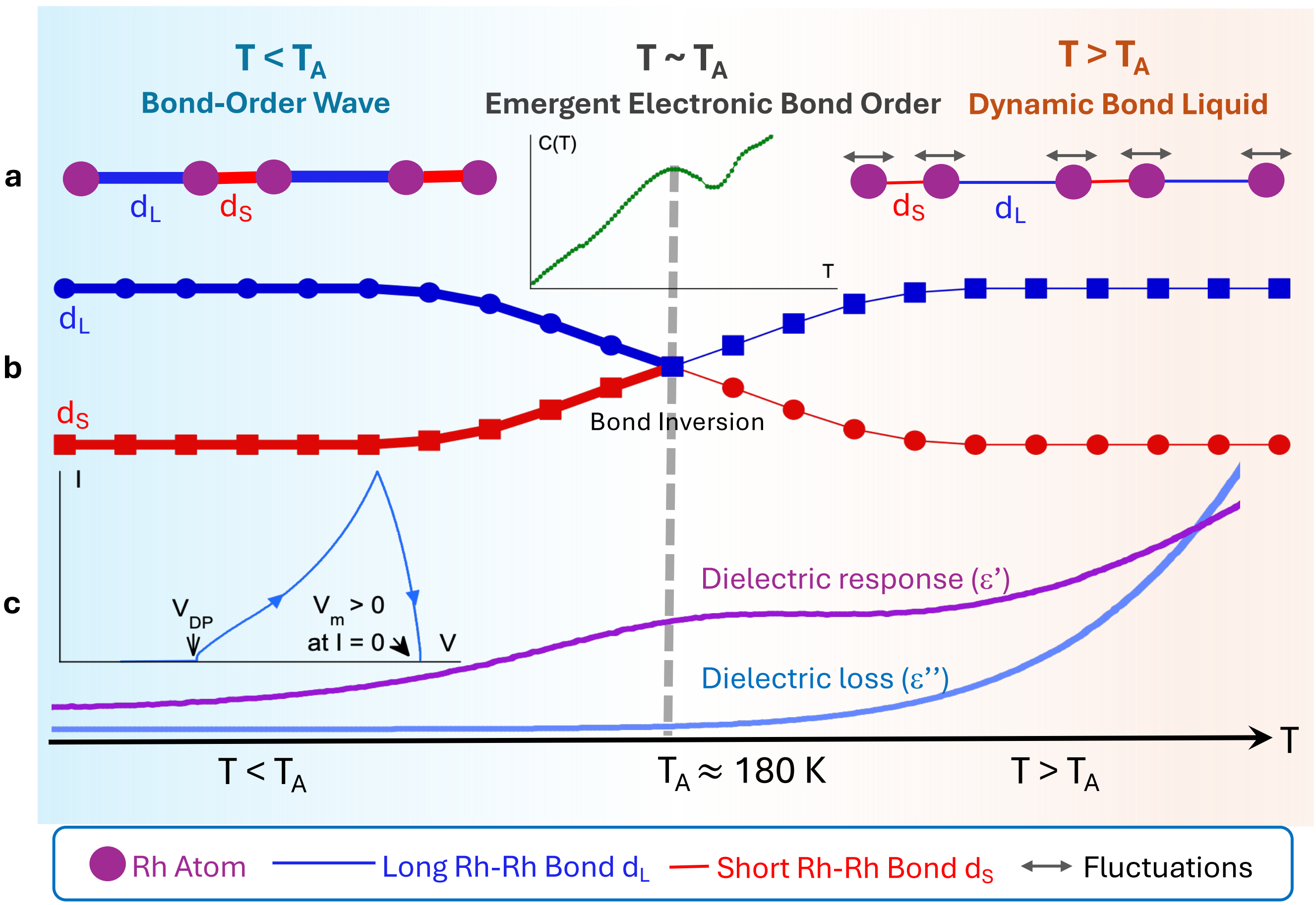

T < T_A
Bond-Order Wave
T ~ T_A
Emergent Electronic Bond Order
T > T_A
Dynamic Bond Liquid
a
b
c
d_L
d_S
C(T)
T
Bond Inversion
I
V_DP
V_m > 0
at I = 0
V
Dielectric response (ε')
Dielectric loss (ε'')
T
T < T_A
T_A ≈ 180 K
T > T_A
Rh Atom
Long Rh-Rh Bond d_L
Short Rh-Rh Bond d_S
Fluctuations

Figure 5

SUPPLEMENTAL INFORMATION

# Emergent Electronic Bond-Order Wave in a Quasi-One-Dimensional Chain

Honghao Wang[1,2], Tristan R. Cao[1], Pedro Schlottmann[3], Gang Cao[1,2*]

[1]Department of Physics, University of Colorado at Boulder, Boulder, CO 80309, USA

[2]Materials Science and Engineering, University of Colorado at Boulder, Boulder, CO 80309, USA

[3]Department of Physics, Florida State University, Tallahassee, FL 32306, USA

## 1. Experimental

Single crystals of $Ba_9Rh_8O_{24}$ were grown using a flux method. Measurements of crystal structures were performed using a Bruker Quest ECO single-crystal diffractometer with an Oxford Cryosystem providing sample temperature environments ranging from 80 K to 400 K. Chemical analyses of the samples were performed using a combination of a Hitachi MT3030 Scanning Electron Microscope and an Oxford Energy Dispersive X-Ray Spectroscopy (EDX). The measurements of the electrical resistivity, I-V characteristics and heat capacity were carried out using a Quantum Design (QD) Dynacool PPMS system having a 14-Tesla magnet and a set of external Keithley meters that provides current source and measures voltage with a high precision. Note that the I-V curves are current-driven. The contact resistance was measured to be on the order of 1 Ω, and these measurements were made using standard four-probe configurations, with current and voltage leads separated to eliminate contact contributions to voltage measurements. The measurements of dielectric constant were performed using a QuadTech 7600 LCR meter and a home-made probe suitable for the PPMS.

$Ba_9Rh_8O_{24}$ exhibits unusually slow relaxation dynamics, with a characteristic relaxation time of approximately 30-35 s. To minimize nonequilibrium effects and allow the system to approach thermal equilibrium during measurements, we employed a deliberately slow cooling rate of 0.3 K/min, corresponding to a temperature change of only approximately 0.15–0.18 K over one relaxation time. This protocol is particularly important for a system exhibiting slow collective dynamics and history-dependent electrical responses

## 2. Supplemental crystallographic data for $Ba_9Rh_8O_{24}$

**Table S1 Crystal data and structure refinement for $Ba_9Rh_8O_{24}$ at 100 K.**

| | |
|---|---|
| Empirical formula | $Ba_9O_{24}Rh_8$ |
| Formula weight | 2443.34 |
| Temperature/K | 100 |
| Crystal system | trigonal |
| Space group | R-3c (No. 167) |
| a/Å | 10.0443(6) |
| b/Å | 10.0443(6) |
| c/Å | 41.279(4) |
| α/° | 90 |
| β/° | 90 |
| γ/° | 120 |
| Volume/Å³ | 3606.6(5) |
| Z | 6 |
| Radiation | Mo Kα (λ = 0.71073 Å) |
| θ range for data collection/° | 2.541–40.366 |
| Reflections collected | 37269 |
| Independent reflections | 2546 |
| Data/restraints/parameters | 2546/0/44 |
| Goodness-of-fit on F² | 1.125 |
| Final R indexes [I ≥ 2σ(I)] | R1 = 0.0819, wR2 = 0.1422 |

**Table S1 (Cont'd) Fractional Atomic Coordinates (×10⁴) and Equivalent Isotropic Displacement Parameters (Å² ×10³) for $Ba_9Rh_8O_{24}$ at 100 K.** Ueq is defined as 1/3 of the trace of the orthogonalised Uij tensor.

| Atom | x | y | z | U(eq) |
|---|---|---|---|---|
| Ba1 | 6815.0(12) | 6666.67 | 4166.67 | 4.95(17) |
| Ba2 | 6626.2(9) | 6763.7(11) | 6389.3(2) | 9.89(16) |
| Rh3 | 3333.33 | 6666.67 | 6666.67 | 2.9(3) |
| Rh5 | 3333.33 | 6666.67 | 6057.7(3) | 4.0(3) |
| Rh4 | 3333.33 | 6666.67 | 5466.4(4) | 4.9(3) |
| Rh6 | 3333.33 | 6666.67 | 4830.6(5) | 8.8(3) |
| Rh7 | 3333.33 | 6666.67 | 4166.67 | 12.9(5) |
| O1 | 3424(11) | 5127(10) | 6360(2) | 3.6(12) |
| O2 | 3275(11) | 8240(11) | 5774(2) | 5.8(15) |
| O3 | 1738(13) | 6521(12) | 5149(2) | 10.9(17) |
| O4 | 4940(30) | 7090(30) | 4508(5) | 50(5) |

## Table S2 Crystal data and structure refinement for $Ba_9Rh_8O_{24}$ at 150 K.

| | |
|---|---|
| Empirical formula | $Ba_9O_{24}Rh_8$ |
| Formula weight | 2443.34 |
| Temperature/K | 150 |
| Crystal system | trigonal |
| Space group | R-3c (No. 167) |
| a/Å | 10.0500(5) |
| b/Å | 10.0500(5) |
| c/Å | 41.294(3) |
| α/° | 90 |
| β/° | 90 |
| γ/° | 120 |
| Volume/Å$^3$ | 3612.0(5) |
| Z | 6 |
| Radiation | Mo Kα (λ = 0.71073 Å) |
| θ range for data collection/° | 2.960–40.346 |
| Reflections collected | 53261 |
| Independent reflections | 2549 |
| Data/restraints/parameters | 2549/0/47 |
| Goodness-of-fit on $F^2$ | 1.043 |
| Final R indexes [I ≥ 2σ(I)] | R1 = 0.0707, wR2 = 0.1338 |

## Table S2 (Cont'd) Fractional Atomic Coordinates ($\times 10^4$) and Equivalent Isotropic Displacement Parameters (Å$^2$ $\times 10^3$) for $Ba_9Rh_8O_{24}$ at 150 K. Ueq is defined as 1/3 of the trace of the orthogonalised Uij tensor.

| Atom | x | y | z | U(eq) |
|---|---|---|---|---|
| Ba1 | 3333.33 | 3155.2(8) | 4166.67 | 4.21(13) |
| Ba2 | 6611.1(8) | 9891.5(7) | 6388.0(2) | 11.81(14) |
| Rh3 | 3333.33 | 6666.67 | 6666.67 | 4.1(3) |
| Rh5 | 3333.33 | 6666.67 | 6058.3(3) | 3.3(2) |
| Rh4 | 3333.33 | 6666.67 | 5465.7(3) | 5.6(2) |
| Rh6 | 3333.33 | 6666.67 | 4832.6(4) | 9.5(2) |
| Rh7 | 3333.33 | 6666.67 | 4166.67 | 12.5(4) |
| O1 | 4936(8) | 6704(9) | 6364.9(18) | 4.9(9) |
| O2 | 1685(9) | 6555(9) | 5768.6(18) | 7.4(12) |
| O3 | 4907(11) | 6764(11) | 5155(2) | 15.1(16) |
| O4 | 3750(30) | 5470(20) | 4510(3) | 69(7) |

**Table S3 Crystal data and structure refinement for $Ba_9Rh_8O_{24}$ at 180 K.**

| | |
|---|---|
| Empirical formula | $Ba_9O_{24}Rh_8$ |
| Formula weight | 2443.34 |
| Temperature/K | 180 |
| Crystal system | trigonal |
| Space group | R-3c (No. 167) |
| a/Å | 10.0522(5) |
| b/Å | 10.0522(5) |
| c/Å | 41.309(3) |
| α/° | 90 |
| β/° | 90 |
| γ/° | 120 |
| Volume/Å³ | 3614.9(4) |
| Z | 6 |
| Radiation | Mo Kα (λ = 0.71073 Å) |
| θ range for data collection/° | 2.539–40.380 |
| Reflections collected | 62338 |
| Independent reflections | 2558 |
| Data/restraints/parameters | 2558/0/50 |
| Goodness-of-fit on $F^2$ | 1.024 |
| Final R indexes [I ≥ 2σ(I)] | R1 = 0.0719, wR2 = 0.1321 |

**Table S3 (Cont'd) Fractional Atomic Coordinates (×10⁴) and Equivalent Isotropic Displacement Parameters (Å² ×10³) for $Ba_9Rh_8O_{24}$ at 180 K. Ueq is defined as 1/3 of the trace of the orthogonalised Uij tensor.**

| Atom | x | y | z | U(eq) |
|---|---|---|---|---|
| Ba1 | 3333.33 | 3161.0(11) | 4166.67 | 5.76(16) |
| Ba2 | 107.8(9) | 6719.3(10) | 6387.7(2) | 14.17(18) |
| Rh3 | 3333.33 | 6666.67 | 6666.67 | 4.4(3) |
| Rh5 | 3333.33 | 6666.67 | 6058.3(3) | 5.7(3) |
| Rh4 | 3333.33 | 6666.67 | 5464.5(4) | 6.2(3) |
| Rh6 | 3333.33 | 6666.67 | 4832.5(5) | 12.3(3) |
| Rh7 | 3333.33 | 6666.67 | 4166.67 | 14.2(5) |
| O1 | 1768(10) | 5067(9) | 6369(2) | 6.5(11) |
| O2 | 1685(11) | 6572(11) | 5769(2) | 9.6(16) |
| O3 | 1832(11) | 5082(12) | 5156(2) | 12.3(16) |
| O4 | 3680(20) | 5440(20) | 4512(5) | 58(5) |

**Table S4 Crystal data and structure refinement for $Ba_9Rh_8O_{24}$ at 200 K.**

| | |
|---|---|
| Empirical formula | $Ba_9O_{24}Rh_8$ |
| Formula weight | 2443.34 |
| Temperature/K | 200 |
| Crystal system | trigonal |
| Space group | R-3c (No. 167) |
| a/Å | 10.0500(4) |
| b/Å | 10.0500(4) |
| c/Å | 41.294(3) |
| α/° | 90 |
| β/° | 90 |
| γ/° | 120 |
| Volume/Å$^3$ | 3612.0(4) |
| Z | 6 |
| Radiation | Mo Kα (λ = 0.71073 Å) |
| θ range for data collection/° | 2.540–38.211 |
| Reflections collected | 51616 |
| Independent reflections | 2224 |
| Data/restraints/parameters | 2224/0/48 |
| Goodness-of-fit on F$^2$ | 1.102 |
| Final R indexes [I ≥ 2σ(I)] | R1 = 0.0666, wR2 = 0.0924 |

**Table S4 (Cont'd) Fractional Atomic Coordinates (×10$^4$) and Equivalent Isotropic Displacement Parameters (Å$^2$ ×10$^3$) for $Ba_9Rh_8O_{24}$ at 200 K. Ueq is defined as 1/3 of the trace of the orthogonalised Uij tensor.**

| Atom | x | y | z | U(eq) |
|---|---|---|---|---|
| Ba1 | 3333.33 | 3146.5(8) | 4166.67 | 3.10(13) |
| Ba2 | 6612.7(8) | 9893.5(7) | 6387.5(2) | 11.14(14) |
| Rh3 | 3333.33 | 6666.67 | 6666.67 | 3.0(3) |
| Rh5 | 3333.33 | 6666.67 | 6058.7(3) | 2.7(2) |
| Rh4 | 3333.33 | 6666.67 | 5465.2(3) | 4.7(2) |
| Rh6 | 3333.33 | 6666.67 | 4832.5(4) | 7.1(2) |
| Rh7 | 3333.33 | 6666.67 | 4166.67 | 10.2(4) |
| O1 | 4937(8) | 6704(10) | 6364(2) | 5.5(10) |
| O2 | 1681(9) | 6558(9) | 5769.9(17) | 5.3(12) |
| O3 | 4906(12) | 6779(11) | 5154(2) | 14.5(17) |
| O4 | 3780(20) | 5510(20) | 4505(3) | 61(6) |

## Table S5 Crystal data and structure refinement for $Ba_9Rh_8O_{24}$ at 250 K.

| | |
|---|---|
| Empirical formula | $Ba_9O_{24}Rh_8$ |
| Formula weight | 2443.34 |
| Temperature/K | 250 |
| Crystal system | trigonal |
| Space group | R-3c (No. 167) |
| a/Å | 10.0530(6) |
| b/Å | 10.0530(6) |
| c/Å | 41.305(4) |
| α/° | 90 |
| β/° | 90 |
| γ/° | 120 |
| Volume/Å$^3$ | 3615.1(5) |
| Z | 6 |
| Radiation | Mo Kα (λ = 0.71073 Å) |
| θ range for data collection/° | 2.539–40.384 |
| Reflections collected | 56576 |
| Independent reflections | 2565 |
| Data/restraints/parameters | 2565/0/50 |
| Goodness-of-fit on $F^2$ | 1.043 |
| Final R indexes [I ≥ 2σ(I)] | R1 = 0.0713, wR2 = 0.1169 |

## Table S5 (Cont'd) Fractional Atomic Coordinates (×10$^4$) and Equivalent Isotropic Displacement Parameters (Å$^2$ ×10$^3$) for $Ba_9Rh_8O_{24}$ at 250 K. Ueq is defined as 1/3 of the trace of the orthogonalised Uij tensor.

| Atom | x | y | z | U(eq) |
|---|---|---|---|---|
| Ba1 | 3333.33 | 3191.3(11) | 4166.67 | 8.79(19) |
| Ba2 | 3237.3(11) | 3374.1(9) | 6389.4(2) | 14.46(18) |
| Rh3 | 3333.33 | 6666.67 | 6666.67 | 4.3(4) |
| Rh5 | 3333.33 | 6666.67 | 6057.9(3) | 5.9(3) |
| Rh4 | 3333.33 | 6666.67 | 5464.4(4) | 6.3(3) |
| Rh6 | 3333.33 | 6666.67 | 4830.0(5) | 11.4(3) |
| Rh7 | 3333.33 | 6666.67 | 4166.67 | 16.3(5) |
| O1 | 4880(10) | 6599(12) | 6363(2) | 7.1(12) |
| O2 | 3317(11) | 5058(12) | 5772(2) | 9.7(17) |
| O3 | 3444(15) | 8240(12) | 5154(2) | 19(2) |
| O4 | 2980(20) | 5130(30) | 4507(5) | 66(6) |

**Table S6 Crystal data and structure refinement for $Ba_9Rh_8O_{24}$ at 300 K.**

| | |
|---|---|
| Empirical formula | $Ba_9O_{24}Rh_8$ |
| Formula weight | 2443.34 |
| Temperature/K | 300 |
| Crystal system | trigonal |
| Space group | R-3c (No. 167) |
| a/Å | 10.0663(5) |
| b/Å | 10.0663(5) |
| c/Å | 41.348(3) |
| α/° | 90 |
| β/° | 90 |
| γ/° | 120 |
| Volume/Å$^3$ | 3628.5(5) |
| Z | 6 |
| Radiation | Mo Kα (λ = 0.71073 Å) |
| θ range for data collection/° | 2.536–40.192 |
| Reflections collected | 78081 |
| Independent reflections | 2545 |
| Data/restraints/parameters | 2545/0/55 |
| Goodness-of-fit on F$^2$ | 1.035 |
| Final R indexes [I ≥ 2σ(I)] | R1 = 0.0727, wR2 = 0.1314 |

**Table S6 (Cont'd) Fractional Atomic Coordinates (×10$^4$) and Equivalent Isotropic Displacement Parameters (Å$^2$ ×10$^3$) for $Ba_9Rh_8O_{24}$ at 300 K. Ueq is defined as 1/3 of the trace of the orthogonalised Uij tensor.**

| Atom | x | y | z | U(eq) |
|---|---|---|---|---|
| Ba1 | 9843.9(12) | 3177.3(12) | 4166.67 | 12.5(2) |
| Ba2 | 6625.5(9) | 6753.8(12) | 6392.3(2) | 15.20(19) |
| Rh3 | 3333.33 | 6666.67 | 6666.67 | 4.2(3) |
| Rh5 | 3333.33 | 6666.67 | 6057.7(3) | 8.4(3) |
| Rh4 | 3333.33 | 6666.67 | 5464.6(3) | 6.9(3) |
| Rh6 | 3333.33 | 6666.67 | 4827.7(5) | 13.5(3) |
| Rh7 | 3333.33 | 6666.67 | 4166.67 | 20.1(5) |
| O1 | 3435(10) | 5133(9) | 6369(2) | 5.5(11) |
| O2 | 3270(12) | 8229(13) | 5761(2) | 13.7(19) |
| O3 | 1749(11) | 6562(12) | 5158(2) | 14.9(19) |
| O4 | 2127(20) | 5103(20) | 4503(4) | 58(5) |

## 3. Supplemental heat capacity, thermal conductivity and I-V data

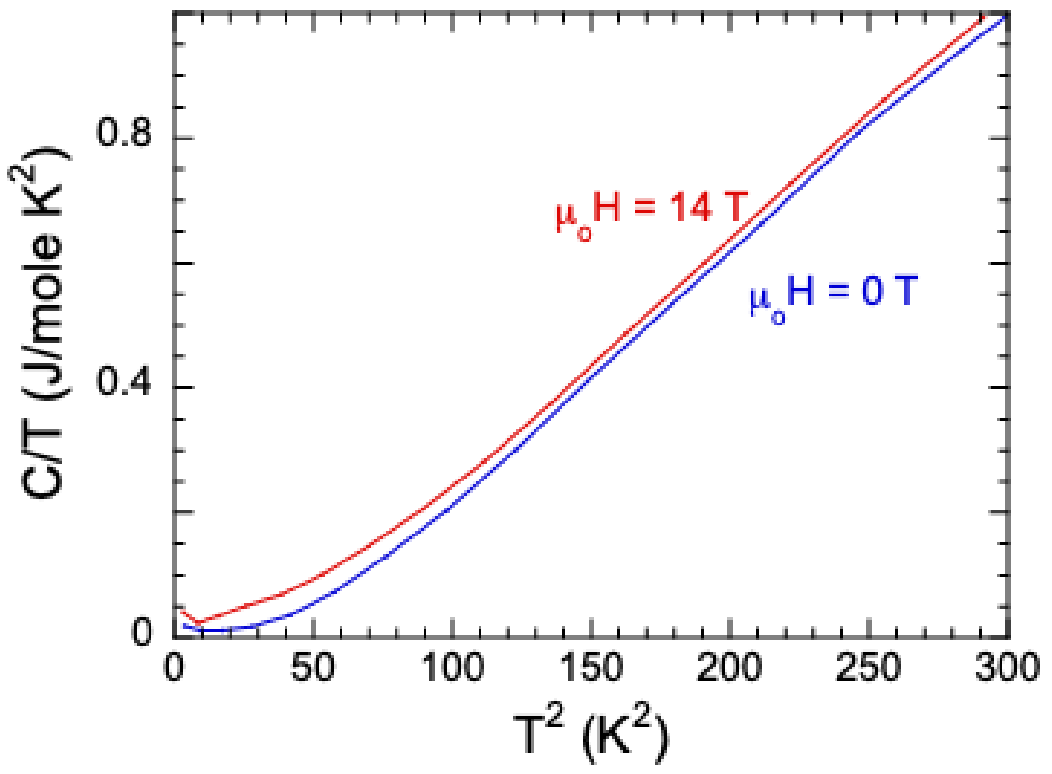

***SFig.1. Heat capacity C at low temperatures under magnetic fields, 0 and 14 T.*** *The data in the C/T vs $T^2$ plot changes only slightly at 14 T. Note the nonlinear behavior below T = 10 K ($T^2$ = 100 $K^2$).*

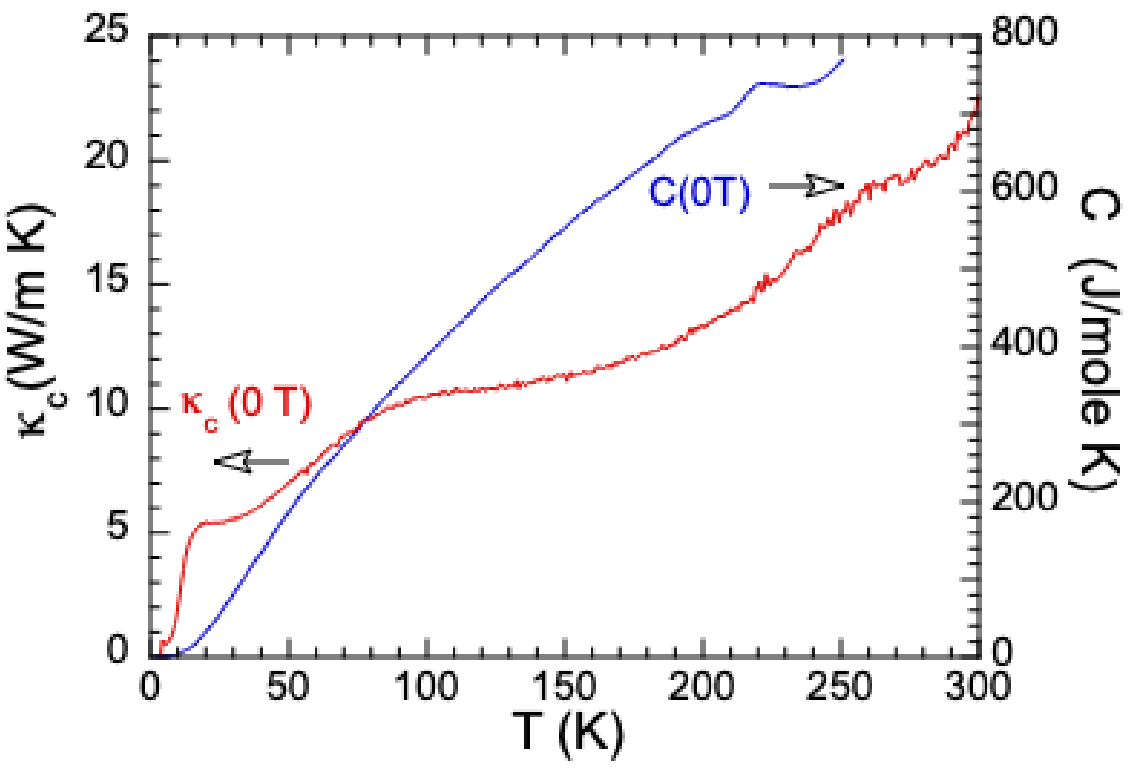

***SFig.2. Temperature dependence of the c-axis thermal conductivity $\kappa_c(T)$ (red, left axis) and heat capacity C(T) (blue, right axis) of $(Ba_{1-x}La_x)_9Rh_8O_{24}$ with x = 0.02 at 0T.*** *Unlike the linear $\kappa_c(T)$ of the pristine compound, the La-substituted sample exhibits pronounced curvature and a rapid increase at higher temperatures, while C(T) varies comparatively smoothly. The contrasting temperature dependences suggest that thermal transport is particularly sensitive to changes in scattering processes induced by modest chemical substitution.*

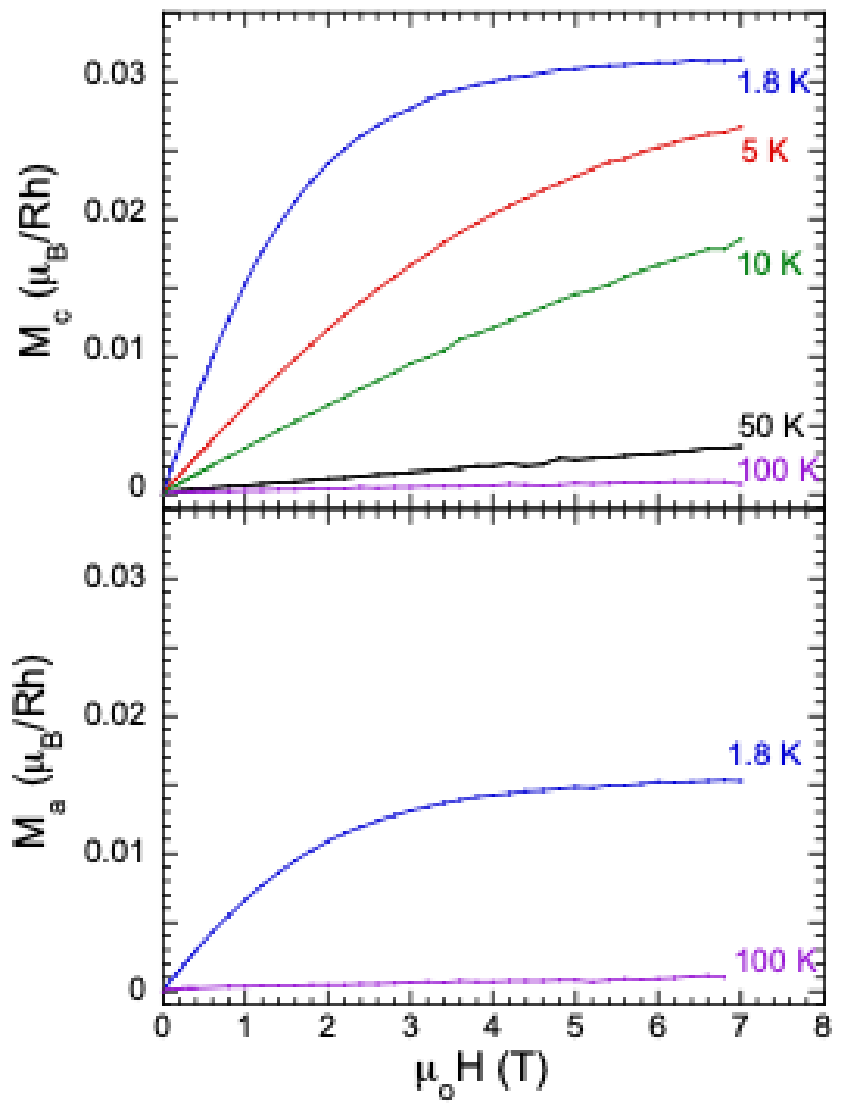

***SFig.3. Upper:*** *c-axis $M_c$ at selected temperatures from 1.8 to 100 K.* ***Lower:*** *Corresponding a-axis Ma at 1.8 and 100 K for the pristine sample.*

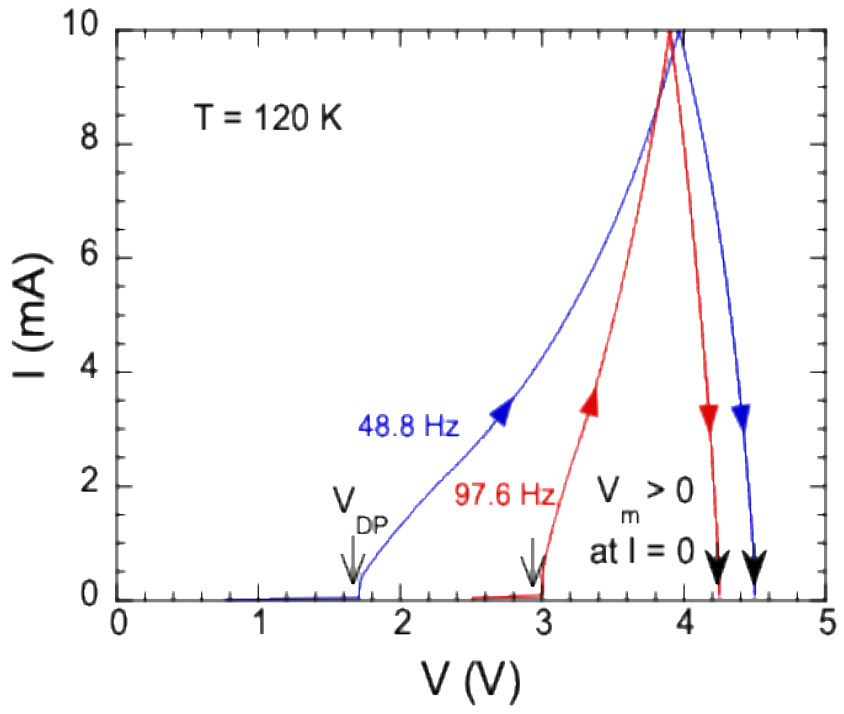

***SFig.4. I-V characteristics at 120 K for f = 48.8 and 97.6 Hz.*** *The driving f shifts depinning voltage $V_{DP}$ (thin arrows) and the remnant voltage $V_m$ (thick arrows) below $T_A$.*

## 4. Transition $T_A$

We note that the anomalies in the lattice, heat capacity, magnetization, and dielectric response, discussed below, occur at slightly different temperatures, but within the same regime. Such small offsets are expected for a transition involving strongly coupled structural and electronic degrees of freedom, because different probes couple to distinct components and timescales of the order parameter. Their close correspondence defines a common transition regime near $T_A \approx 180$ K associated with the collective reorganization of the Rh-Rh bond network.

## 5. Joule heating unlikely to account for the nonlinear and memristive response

Most directly, the clockwise hysteresis leaves a finite remanent voltage $V_m$ when the current returns to zero (Fig.4 and SFig.2), demonstrating that the electrical state retains memory after the instantaneous Joule power $I^2R$ has vanished. In addition, on the return branch V can remain nearly constant or even increase as I decreases, which is difficult to reconcile with a simple monotonic temperature rise and recovery. The clockwise hysteresis also becomes more pronounced with increasing driving frequency and, above $T_A$, appears only above a characteristic frequency, whereas a purely thermal mechanism would generally be expected to become more effective for slower sweeps that allow more time for heat accumulation and equilibration. Finally, the strong evolution of the nonlinear response through $T_A$ and its correlation with the structural, thermodynamic, and dielectric signatures of the BOW point to an intrinsic collective origin. These observations are instead consistent with current-driven reorganization of the bond-ordered state into metastable configurations whose relaxation is slow compared with the electrical driving timescale.